\pdfoutput=1
\documentclass[aps,prb, twocolumn, groupedaddress, amssymb, amsmath,nofootinbib]{revtex4-2}
\usepackage{bm}
\usepackage{graphicx}
\usepackage{float}
\usepackage{todonotes}
\usepackage{physics}

\begin{document}

\title{Effect of Van Hove singularities on Shiba states in two-dimensional $s$-wave superconductors}

\author{Mateo Uldemolins}
\author{Andrej Mesaros}
\author{Pascal Simon}
\email{pascal.simon@universite-paris-saclay.fr}
\affiliation{Universit\'e Paris-Saclay, CNRS, Laboratoire de Physique des Solides, 91405, Orsay, France}

\date{\today}

\begin{abstract}
Magnetic impurities in a superconductor induce Yu-Shiba-Rusinov (YSR) states inside the superconducting gap, whose energy depends on the strength of the coupling to the impurity and on the density of states (DOS) at the Fermi level. We consider DOS exhibiting a logarithmic or a power-law divergence at the Fermi level due to Van Hove singularities (VHS) and high-order Van Hove singularities (HOVHS), respectively. We find that the energy of the YSR states has the same functional form as in the constant DOS scenario, with the effect of the singularity being an enhancement of the effective coupling constants. In particular, the critical magnetic coupling strength at which the Shiba transition occurs is always lowered by a factor $1/\rho(\Delta/E_{\mathrm{c}}$), where $\Delta$ is the superconducting gap, $E_{\mathrm{c}}$ is the bandwidth, and $\rho(E)$ is the factor in DOS which diverges at $E=0$ for a VHS or HOVHS. Further, since the critical magnetic coupling is significantly reduced, a new regime becomes accessible where the transition point is controlled by the non-magnetic coupling constant. Interestingly, the slope of the Shiba energy curve at the Shiba transition is independent of impurity parameters and purely reflects the band structure. Additionally, we find that our main conclusions remain valid even when the Fermi level is not precisely tuned to the Van Hove singularity, but instead lies on an energy range of order the superconducting gap. Our results show that tuning a superconducting material towards a VHS or HOVHS enhances the possibilities for engineering YSR states, and for characterizing the superconductor itself. 
\end{abstract}


\maketitle

\section{Introduction}
The appearance of VHS in the DOS as the system is tuned through a Lifshitz transition enhances electronic correlations which in turn can trigger quantum emergent phenomena such as superconductivity \cite{luttinger:sc, dzialoshinskii:sc} or charge-density-wave instability \cite{rice:cdw}. Recent experimental progress showcasing the ability to bring a VHS arbitrarily close to the Fermi level in twisted bilayer graphene (TBG) \cite{luican:vh_tbg, xu:tunable_vh} and in heterostructures \cite{mori:controlling} has reawakened the interest in the field. On the theory side, the notion of VHS was recently extended to HOVHS where the DOS is enhanced from logarithmic to power-law divergence and they were predicted to be realizable by tuning a single parameter in Moir\'e superlattices \cite{yuan:magic, yuan:classification}. Although some progress has been made to characterize HOVHS in interacting systems \cite{isobe:supermetal}, the interplay of HOVHS and magnetic impurities in conventional superconductors remains unknown.

The classical spin of a magnetic impurity embedded in a superconductor is an example of a pair-breaking defect which leads to the formation of localized excitations inside the superconducting energy gap known as Yu-Shiba-Rusinov (YSR) states \cite{Yu1965,shiba:classical_spins, rusinov, sakurai:comments, salkola:magnetic_moments}. Ever since their first observation by Yazdani et al. \cite{yazdani:probing}, the advancements in scanning tunneling spectroscopy (STS) and atomic manipulation techniques have prompted a growing interest in the subject (see \cite{franke:review_shiba} for a recent review).
YSR states have proved to be a useful  tool to probe the orbital properties of the magnetic adsorbate \cite{franke:orbital, choi:magnetic_ordering, Ast:2018, verdu:molecular_spin}, and recently, the advent of STS functionalized tips has opened a new avenue to study spin-dependent phenomena \cite{huang2020:tunneling, huang2020:spin_dependent, schneider2021:spin_polarization, villas:tunneling}. Simultaneously many works have proposed  YSR as building blocks to engineer exotic topological states of matter \cite{Nakosai2013,NP2013,pientka:topo,Braunecker2013,Klinovaja2013,Vazifeh2013,Kim2014,Heimes2014,Li2014,brydon:topo_chain,Rontynen2015,Braunecker2015,Rontynen2016,Schecter2016,Christen2016,Hoffman2016,Li2016b} with experimental spectral signatures consistent with topological superconductivity displaying Majorana zero modes \cite{nadj-perge:majorana_chain,Ruby2015, Pawlak2016,Yazdani2017,Jeon2017,palacio-morales:majorana_chain} (see \cite{Yazdani2021} for a recent review and refs. therein). Both endeavors require the YSR states to have a large spatial extent as well as a strong coupling to the substrate. While past observations indicate that quasi two-dimensional materials constitute an ideal playground to achieve the former \cite{menard:coherent, wiesendanger:focusing}, a quantitative experimental connection between the YSR state energy and the impurity-substrate hybridization has been only analyzed very recently \cite{franke:tuning,Ast2020} and deserves further study.

In the present work, motivated by the recent discovery of superconductivity in TBG \cite{jarillo:sc_tbg}, the quest to find YSR states in graphene \cite{lado:unconventiona_ysr, rio:ysr_graphene, lado:ysr_tbg}, and the promising pathway to VHS tunability in these materials, we explore the influence of conventional and high-order VHS on YSR states. We find that tuning the Fermi level to a Van Hove singularity is a powerful way to enhance the coupling of YSR states to the substrate, while at the same time, it provides a possibility to extract information from the DOS singularity itself.

Our paper is organized as follows. In Sec.~\ref{sec:model}, we present our model Hamiltonian for the Shiba impurity coupled to a superconductor with a Fermi level tuned near a VHS or HOVHS. Sec.~\ref{sec:results} contains our analytical and numerical results for both types of Van Hove singularities. Finally, in Sec.~\ref{sec:conclusion} we provide a short conclusion and discussion of our results. The effect of a small perturbation in the chemical potential away from the singularity is discussed in App.~\ref{app:mu}, and some technical details are presented in App.~\ref{app:integrals}.

\section{Model Hamiltonian}
\label{sec:model}
We consider a pointlike, isotropic, magnetic impurity on a two-dimensional, $s$-wave superconductor. The Bogoliubov-de Gennes Hamiltonian (BdG) of the system in the Nambu basis $\Psi = (\psi_{\uparrow}, \psi_{\downarrow}, \psi_{\downarrow}^\dagger, -\psi_{\uparrow}^\dagger)^T$ reads
\begin{equation}
\mathcal{H} = \xi_{\bm{k}} \tau_z + \Delta \tau_x + (K \tau_z - J \sigma_z) \delta(\bm{r} - \bm{r}_{\mathrm{imp}}),
\label{eq:bdg_ham}
\end{equation}
where $\bm{k}$ and $\bm{r}$ designate the electron's momentum and position, $\xi_{\bm{k}}$ is the energy dispersion of the electrons in the superconducting substrate, $\Delta$ is the superconducting gap, $K$ is the amplitude of the non-magnetic scattering potential and $J\equiv JS/2$ denotes the coupling strength between the electrons and the magnetic impurity with classical spin $S$ at $\bm{r}_{\mathrm{imp}}$, and Pauli matrices $\tau_i$ and $\sigma_i$ act on particle-hole and spin space, respectively. In the present study we disregard any quantum effects associated with the magnetic impurity (e.g. Kondo screening), and comment on the validity of the model in the discussion section.

In this work we focus on three expressions for the energy dispersion $\xi_{\bm{k}}$ defined in the continuum, since the information relevant for the Shiba state comes from the vicinity of the Fermi momentum, and thereby the three energy dispersions are representative of three broad classes of systems: ones with constant DOS at the Fermi energy, ones with a classical VHS, and ones belonging to a family of HOVHS labeled by a parameter $\nu$, as we detail now. Firstly, owing to the toroidal topology of the Brillouin zone, the energy dispersion of two-dimensional systems is always endowed with at least two saddle points, defined as
\begin{equation}
 \bm{\nabla_k} \xi_{\bm{k}}= 0, \quad \det D < 0,
\end{equation}
where $D$ the Hessian matrix of $\xi_{\bm{k}}$. First discussed in the context of phonons \cite{van_hove} and later applied to electronic systems, it is well known that such saddle points yield a logarithmic singularity in the DOS at the corresponding energy, henceforth denoted conventional Van Hove singularity:
\begin{equation}
\label{eq:dos_cvh}
\rho(\xi) = \frac{1}{2E_{\mathrm{c}}} \log\left(\left|\frac{E_\mathrm{c}}{\xi}\right|\right),
\end{equation}
where $\xi$ is the energy measured from the Fermi level. Here the prefactor stems from imposing the normalization condition $1 = \int_{-E_{\mathrm{c}}}^{E_{\mathrm{c}}} \rho(\xi) d\xi$, and $E_\mathrm{c}$ is an energy cutoff which is introduced to delimit the energy range over which Eq.~\eqref{eq:dos_cvh} is a good description of the system's DOS. In our calculations we always assume that $E_{\mathrm{c}}$ is the largest energy scale, of the order of the bandwidth. Next, for systems with constant density of states at the Fermi energy, we use the constant
\begin{equation}
\label{eq:dos_const}
\rho(\xi) = \frac{1}{2E_{\mathrm{c}}},
\end{equation}
which is conveniently normalized with the use of the same bandwidth. Finally, in recent works the notion of Van Hove singularity was extended to high order Van Hove singularities  \cite{yuan:classification}, where degenerate saddle points, i.e. fulfilling $\det D = 0$, lead to a power-law divergent DOS, 
\begin{equation}
\label{eq:dos_hovh}
\rho(\xi) = \frac{\nu+1}{\eta+1}\frac{1}{E_{\mathrm{c}}^{\nu+1}} \; |\xi|^{\nu} \begin{cases} 
      \eta \quad \mathrm{if} \quad \xi < 0, \\
      1 \quad \mathrm{if} \quad \xi > 0.\end{cases}
\end{equation}
Here $-1 < \nu < 0$, $\nu \in \mathbb{Q}$, is the parameter labeling the HOVHS, the $E_{\mathrm{c}}$ has the same meaning as above, and $\eta  \equiv \frac{\rho(-|\xi|)}{\rho(|\xi|)}$ is the particle-hole asymmetry ratio. Note that we include the case of particle-hole symmetric HOVHS at $\eta=1$, and that the previous two dispersions are also symmetric. The exponent $\nu$ can take an infinite number of rational values in its range, each value having a corresponding $\eta$ value; a set of possibilities is tabulated in Ref. \cite{yuan:classification}.

Having defined the models, we now calculate the energy of the Shiba states looking for in-gap solutions of the Schr\"odinger equation in particle-hole space.



\section{Results and discussion}
\label{sec:results}
\subsection{Conventional Van Hove singularity}

\begin{figure}
\centering
\includegraphics[width=\columnwidth]{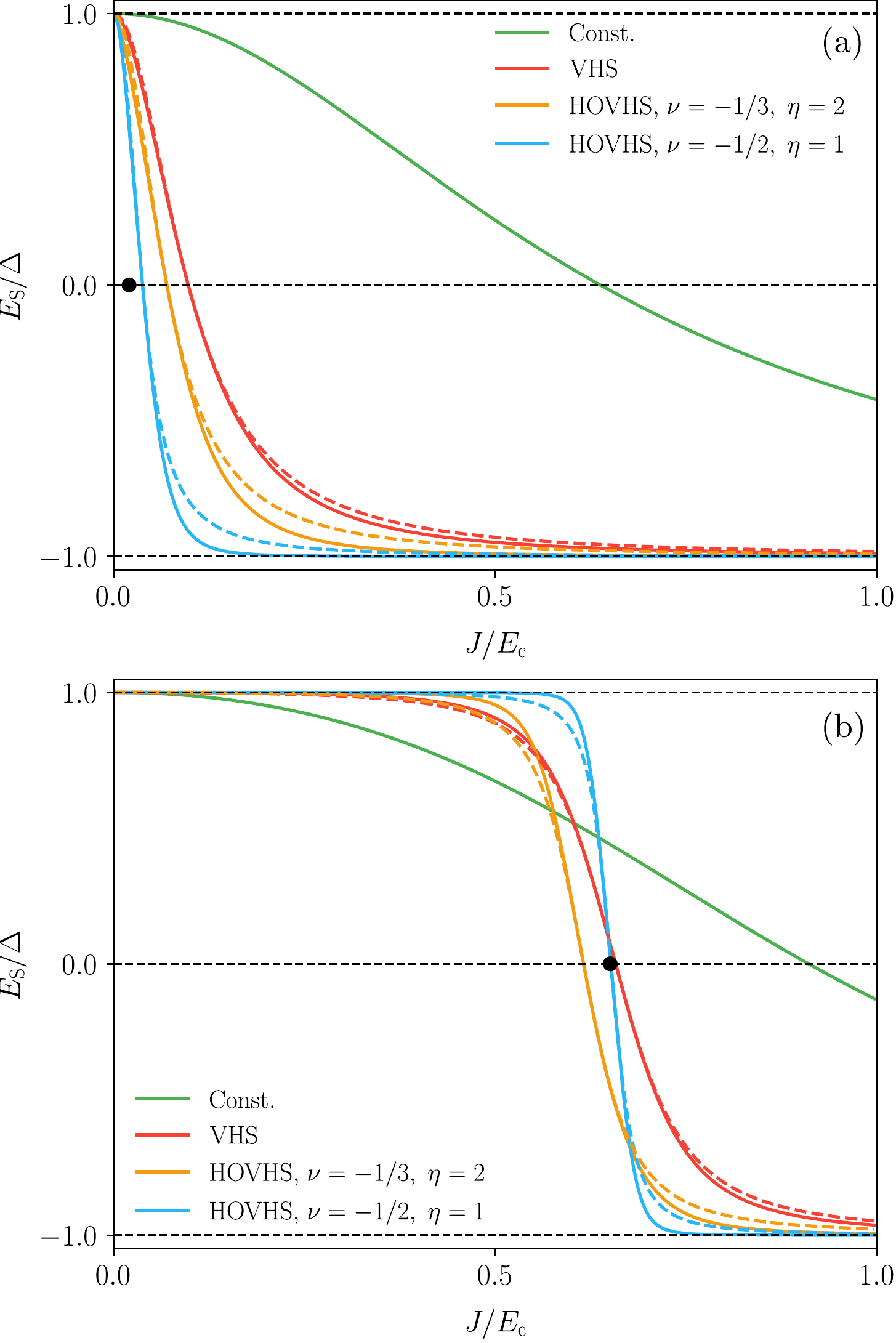}%
\caption{Positive branch of the Shiba energy, Eqs.~\eqref{eq:shiba_energy} and \eqref{eq:shiba_energy_full}, as a function of the magnetic coupling strength $J$ for a constant (green), logarithmically-divergent (red), power-law divergent with $\nu=-1/3$, $\eta = 2$ (orange) and power-law divergent with $\nu=-1/2$, $\eta=1$ (blue) DOS. The black dot indicates the value of $K/E_{\mathrm{c}}$. In the small $K$ regime (a), $J_{\mathrm{c}}$ depends strongly on the underlying DOS, while for large $K$ regime (b), $J_{\mathrm{c}} \sim K$. The solid lines indicate the numerical solution of the self-consistent equations \eqref{eq:shiba_energy} and \eqref{eq:shiba_energy_full} while the dashed lines represent the zeroth-order approximation in $E_{\mathrm{S}}$.}
\label{fig:spec}
\end{figure}

A logarithmic divergence of the DOS at the Fermi level as in Eq.~\eqref{eq:dos_cvh}, yields the following  self-consistent expression for the Shiba energy (see Appendix \ref{app:integrals} for details),
\begin{equation}
\label{eq:shiba_energy}
E_{\mathrm{S}} = \pm \Delta \frac{1-\widetilde{J}^2 + \widetilde{K}^2}{\sqrt{4\widetilde{J}^2 + \left(1-\widetilde{J}^2+\widetilde{K}^2\right)^2}},
\end{equation}
where the existence of two symmetric eigenvalues stems from the particle-hole symmetry constraint of the BdG Hamiltonian, and $\widetilde{J} = f_{\mathrm{vh}}(\Delta, E_{\mathrm{S}}) J$ and $\widetilde{K} = f_{\mathrm{vh}}(\Delta, E_{\mathrm{S}})K$ are effective coupling energies with
\begin{equation}
 f_{\mathrm{vh}}(\Delta, E_{\mathrm{S}}) = \frac{\pi}{2 E_{\mathrm{c}}} \log\left(\frac{E_\mathrm{c}}{\sqrt{\Delta^2 - E_{\text{S}}^2}}\right),
\end{equation}
a renormalizing factor proportional to the density of states in Eq.~\eqref{eq:dos_cvh}.  An expression for $E_{\mathrm{S}}$ of the classical VHS in the strong-coupling limit $E_{\mathrm{S}}\ll\Delta$, and for vanishing $K$ was first obtained in Ref. \cite{tifrea}. Remarkably our expression for the Shiba energy in Eq.~\eqref{eq:shiba_energy} has the same functional form as the one for a constant DOS, namely taking the dispersion in Eq.~\eqref{eq:dos_const} one obtains the standard Shiba energy \cite{rusinov}, i.e. Eq.~\eqref{eq:shiba_energy} with
\begin{equation}
f_{\mathrm{const}} = \frac{\pi}{2 E_{\mathrm{c}}}.
\end{equation}
An important consequence is that a Van Hove singularity in the DOS at the Fermi level enhances the effective coupling energies in the Shiba problem, since $f_{\mathrm{const}} < f_{\mathrm{vh}}$. In the presence of a magnetic impurity, localizing an electron entails an energy trade-off due to the exchange interaction, therefore the energy of the in-gap excitation $E_{\mathrm{S}}$ decreases as $J$ increases (see Fig.~\ref{fig:spec}). This competition triggers the Shiba quantum phase transition at $E_{\mathrm{S}} = 0$, where the in-gap state becomes occupied and it turns into the new superconducting ground state \cite{sakurai:comments, salkola:magnetic_moments}. The critical magnetic coupling $J_{\mathrm{c}}$ prompting the phase transition reads
\begin{equation}
  \label{eq:jc_cvh}
 J^j_{\mathrm{c}} = \sqrt{K^2 + \frac{1}{f_j(\Delta, 0)^2}},
\end{equation}
where $j\in\{\textrm{const},\textrm{vh}\}$ labels the chosen dispersion. In all our calculations we assume $K \ll E_{\mathrm{c}}$, therefore the relation $f_{\mathrm{const}}(\Delta,0) < f_{\mathrm{vh}}(\Delta,0)$ implies that a Van Hove singularity in the DOS at the Fermi level reduces $J_{\mathrm{c}}$ with respect to its value with a constant DOS. This remains true even when Fermi level is not exactly tuned to the Van Hove singularity but within an interval of the order of the superconducting gap instead, although the expression for $J_{\mathrm{c}}^{\mathrm{vh}}$ becomes slightly more complicated (see App.~\ref{app:mu} for details).

The DOS enhancement at the VHS enables one to access a new regime of the impurity system. Namely, the theory is always applicable in a ``small $K$'' regime defined by $K \lesssim \frac{1}{f_j(\Delta,0)}$, which in the case of the constant DOS reiterates the theoretical condition $K\lesssim E_\mathrm{c}$. This regime in the VHS case leads to the critical magnetic coupling $J_{\mathrm{c}}$ reduction due to the larger $f_j(\Delta,0)$ [Fig. \ref{fig:spec} (a)]. However, for a VHS system, we can consider a ``large $K$'' regime, where $K\gg\frac{1}{f_{\mathrm{vh}}(\Delta,0)}$ without contradicting any assumption of the model, and then the $J^j_{\mathrm{c}} \sim K$ would become independent of the DOS [see Fig. \ref{fig:spec} (b)].

We note that the VHS introduces a dependence of the $J_\mathrm{c}$ on the superconducting energy gap $\Delta$, in contrast to the standard Shiba impurity for constant DOS. Specifically it implies that lowering the superconducting gap enhances the impurity coupling, which may motivate a search for systems with optimal size of superconducting gap given the experimental energy resolution.

The general form of the critical magnetic coupling $J^j_{\mathrm{c}}$ reveals a quantity that is independent on the impurity parameters $J,K$ and instead characterizes the DOS itself. Namely, the slope of the $E_{\mathrm{S}}(J)$ curve at the parity-switching point only depends on the DOS and $\Delta$,
\begin{equation}
 \dv{E_{\mathrm{S}}}{J}\Bigr|_{J=J_{\mathrm{c}}} = - \Delta f_j(\Delta,0).
\end{equation}

This is also the case when the chemical potential $\mu$ lies on a $[-\Delta, \Delta]$ interval around the singularity. As shown in App.~\ref{app:mu}, a small displacement of the chemical potential yields a $\mu$-dependent slope but its relative change as $\mu$ varies along the aforementioned interval is of the order of 1\%. In Fig.~\ref{fig:slopes}, we illustrate the unchanging nature of the slope which could be measured in experiments by varying the coupling constant across the transition as it was recently reported \cite{franke:tuning, fan:majorana_vortex, malavolti:tunable}. It could be possible to obtain an estimate of the renormalization parameter, and subsequently to extract information about the nature of the DOS divergence.

\subsection{Higher order Van Hove singularity}
\label{subsec:hovh}
In the case of a power-law divergent DOS at the Fermi level as in Eq.~\eqref{eq:dos_hovh}, the self-consistent expression for the Shiba energy reads,
\begin{widetext}
\begin{equation}
\label{eq:shiba_energy_full}
 E_{\mathrm{S}} = \pm \Delta \frac{1-2f_{2,\nu}K + (f_{1,\nu}^2+f_{2,\nu}^2)(K^2-J^2)}{\sqrt{4f_{1,\nu}^2J^2 + (1-2f_{2,\nu}K + (f_{1,\nu}^2+f_{2,\nu}^2)(K^2-J^2))^2}},
\end{equation}
\end{widetext}
where the renormalization parameters
\begin{subequations}
\label{eq:f1_f2_hovh}
\begin{align}
 f_{1,\nu}(\Delta, E_{\mathrm{S}}) &= \frac{\pi}{2 E_{\mathrm{c}}} \frac{1+\nu}{\cos\left(\frac{\pi}{2}\nu\right)} \left(\frac{\sqrt{\Delta^2 + E_{\mathrm{S}}^2}}{E_{\mathrm{c}}} \right)^{\nu},
\\
 f_{2,\nu}(\Delta, E_{\mathrm{S}})  &=  \frac{\pi}{2 E_{\mathrm{c}}} \frac{1-\eta}{1+\eta} \frac{1+\nu}{\sin\left(\frac{\pi}{2}\nu\right)} \left(\frac{\sqrt{\Delta^2 + E_{\mathrm{S}}^2}}{E_{\mathrm{c}}} \right)^{\nu},
\end{align}
\end{subequations}
also inherit the structure of the DOS. The seemingly more complicated appearance of the Shiba-energy equation stems from the particle-hole asymmetry of the DOS. Indeed, if the DOS is symmetric around the Fermi level, i.e. $\eta = 1$, the factor $f_{2,\nu}$ vanishes and the expression for the Shiba energy simplifies to Eq.~\eqref{eq:shiba_energy} where now $\widetilde{J} = f_{1,\nu}(\Delta, E_{\mathrm{S}}) J$ and $\widetilde{K} = f_{1,\nu}(\Delta, E_{\mathrm{S}})K$. Since $-1 < \nu < 0$, for a sufficiently large energy cutoff $E_{\mathrm{c}}$ we have $f_{1,\nu} > f_{\mathrm{vh}} > f_{\mathrm{const}}$ and we obtain the general result: The effect of a particle-hole symmetric singularity (either a VHS or a HOVHS) in the DOS at the Fermi level is to enhance the effective coupling between the impurity and the electrons in the host.

The critical magnetic coupling for the Shiba transition in case of HOVHS takes a slightly more complicated form,  
\begin{equation}
 \label{eq:jc_hovh}
 J_{\mathrm{c}}^{\mathrm{hovh},\nu} = \sqrt{K^2 + \frac{1-2 f_{2,\nu}(\Delta, 0) K}{f_{1,\nu}(\Delta, 0)^2 + f_{2,\nu}(\Delta, 0)^2}}.
\end{equation}
For the particle-hole symmetric case, $\eta=1$, this expression has the same form as for the VHS, Eq.~\eqref{eq:jc_cvh}. Therefore, regarding the effect of the non-magnetic scattering $K$ for the particle-hole symmetric cases of HOVHS we conclude the same about two possible regimes (small vs. large $K$) as discussed for the VHS after Eq.~\eqref{eq:jc_cvh}.
For the particle-hole asymmetric case, $\eta\neq1$, we note that for reasonable values of the non-magnetic scattering potential, i.e. $K \ll E_{\mathrm{c}}$, the divergence in the DOS always reduces the critical coupling $J_{\mathrm{c}}$ with respect to its value for a constant DOS. As shown in Fig.~\ref{fig:spec} (a), in the small $K$ regime, the stronger the divergence of the DOS, the smaller $J_{\mathrm{c}}$. We note that the $f_{2,\nu}(\Delta, 0) K$ term stemming from the DOS asymmetry would only become relevant if $|K| \gg \frac{1}{f_{2,\nu}(\Delta,0)} \sim E_{\mathrm{c}}^{1+\nu}$, i.e. in the large $K$ regime, but then the $K^2$ term prevails, therefore $J_{\mathrm{c}}$ is governed by the non-magnetic scattering amplitude $K$ and it does not depend on the details of the underlying DOS [Fig.~\ref{fig:spec} (b)]. We remark that when the Fermi level is tuned within an interval of the order of $\Delta$ around the Van Hove singularity, particle-hole symmetric DOSs ($\eta = 1$) also lead to a Shiba energy and critical $J$ expressions of the form \eqref{eq:shiba_energy_full} and \eqref{eq:jc_hovh} respectively, where the $f_{2,\nu}$ factor stems from the asymmetry induced by the displacement of the chemical potential. Nevertheless, the main features presented here remain the same (see App.~\ref{app:mu} for details). 

\begin{figure}
\centering
\includegraphics[width=\columnwidth]{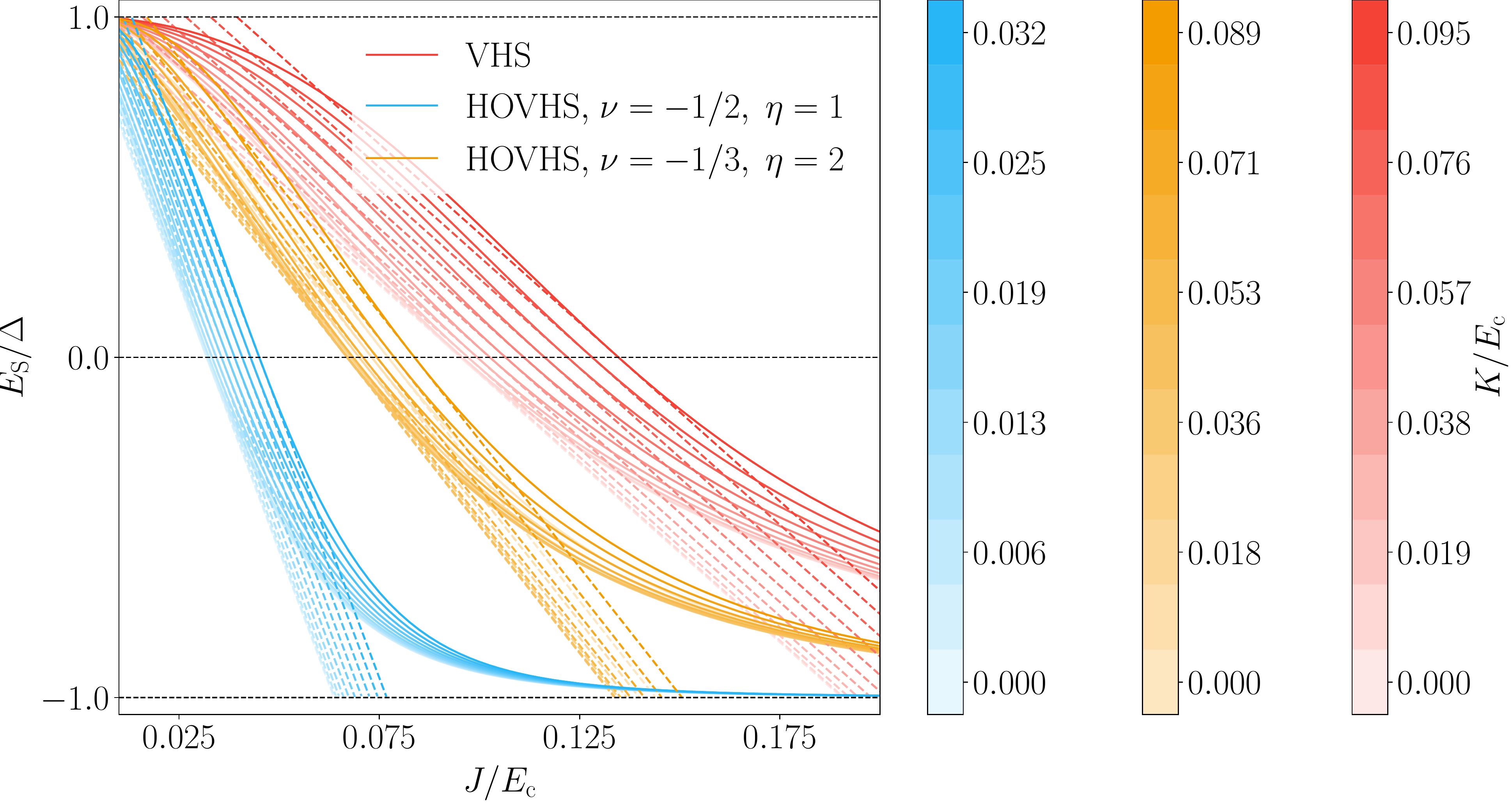}%
\caption{Positive branch of the Shiba energy, Eqs.~\eqref{eq:shiba_energy} and \eqref{eq:shiba_energy_full}, as a function of the magnetic coupling strength $J$ zoomed around the Shiba transition for logarithmically-divergent (red), power-law divergent with $\nu=-1/3$, $\eta=2$ (orange), and power-law divergent with $\nu=-1/2$, $\eta=1$ (blue) DOS. The tangent, dashed lines indicate the slope at the crossing point which remains constant for increasing values of $K$ denoted in the color bars and which becomes steeper for a stronger divergence. Note that if the DOS is asymmetric, the minimum of $J_{\mathrm{c}}$ is not at $K=0$ [cf. blue and orange curves, and Eq.~\eqref{eq:jc_hovh} with and without vanishing $f_{2,\nu}(\Delta, 0)$ factor].}
\label{fig:slopes}
\end{figure}

Interestingly, we again find a quantity that is independent of the impurity parameters, namely the slope of the $E_{\mathrm{S}}(J)$ curve at the Shiba transition point:
\begin{equation}
\label{eq:slope_hovh}
 \dv{E_{\mathrm{S}}}{J}\Bigr|_{J=J_{\mathrm{c}}} = - \Delta \frac{f_{1,\nu}(\Delta, 0)^2 + f_{2,\nu}(\Delta, 0)^2}{f_{1,\nu}(\Delta,0)},
\end{equation}
implying also that the particle-hole asymmetry of the HOVHS divergence would be reflected in Shiba states (see Fig.~\ref{fig:slopes}).

\section{Conclusions}
\label{sec:conclusion}
In summary, we found that tuning the Fermi level to a conventional VHS or HOVHS enhances the coupling between the superconducting substrate and the magnetic impurity, thereby offering a pathway to improve YSR states engineering. These results become particularly significant in the advent of twisted transition metal dichalcogenides and graphene heterostructures which allow to tune the system over large regions of the parameter space. From a practical standpoint one might inquire about the validity of the results in the vicinity of the Van Hove singularity. We explore this matter in detail in Appendix \ref{app:mu}. The key takeaway is that our main conclusions remain valid even when the Fermi level lies on an interval of the order of the smallest energy scale in the problem, i.e. $\mu \in [-\Delta, \Delta]$. If the chemical potential deviates from the Van Hove singularity by an energy of the order of the superconducting gap, essentially the same electronic states are mixed when superconductivity is switched on, hence a similar behavior of the Shiba states is expectable. The robustness against perturbations in the chemical potential strengthens the experimental relevance of these results; nevertheless, it is fair to remark here we only considered an $s$-wave superconductor while electronic instabilities due to a divergent DOS may yield more exotic pairings \cite{yao:topo_vh} or even destroy superconductivity \cite{park:maTTG,hao:maTTG}. Moreover, in this work we treated the spin impurity as a classical degree of freedom. For the conventional case of constant density of states, this approach can be theoretically supported by three arguments: by invoking the limit of a large impurity spin in the Kondo model, by the finding that the Kondo Hamiltonian with strong on-site anisotropy typical of impurity adsorption on a surface yields a sub-gap structure similar to that obtained in the classical limit \cite{zitko:adsorbates}, or by the expectation that typical measurement temperatures are sufficiently higher than the Kondo temperature yet sufficiently lower than the superconducting transition temperature so that the quantum effects of impurity are not pronounced. In our case of a diverging density of states, we expect that the first two arguments still hold, however, theory predicts an enhancement of the Kondo temperature \cite{gogolin1993theory} which might lead to a reduction of the range of experimental validity of the Shiba model compared to the case of constant density of states. An interesting open challenge is to consider the effect of (HO)VHS on a quantum impurity, in a Kondo model or an even more general Anderson impurity model.\\ 

Additionally, we showed that it is possible to extract an impurity-independent quantity from the $E_{\mathrm{S}}-J$ curves, namely the slope of the curve at the Shiba transition. State-of-the-art experiments have shown that it is possible to continuously tune the exchange coupling constant between magnetic molecules \cite{franke:tuning, malavolti:tunable} or magnetic adatoms \cite{fan:majorana_vortex} to the substrate by varying the distance between the impurity and the STM tip, thereby unambiguously identifying the transition point. We propose that this technique could be employed to compare the strength of the divergence in the DOS of different compounds regardless of the nature of the impurities, and even provide an estimate of the divergence law.


\appendix


\section{Perturbation in the chemical potential away from the Van Hove singularity}
\label{app:mu}
In this Appendix we asses the robustness of the previous results to a deviation in the chemical potential away from the Van Hove singularity of the order of the superconducting gap $\Delta$. To that purpose, we assume that the expressions \eqref{eq:dos_cvh} and \eqref{eq:dos_hovh} introduced in the main text remain a good description of the system's DOS and replace the energy dispersion $\xi_{\mathbf{k}}$ in the Hamiltonian \eqref{eq:bdg_ham} by $\xi_{\mathbf{k}} - \mu$, where $\mu$ indicates the chemical potential. To be consistent with the previous assumptions, $E_{\mathrm{c}}$ must stay the largest energy scale of the problem, therefore we can only address the situations where the chemical potential lies on a comparably small range, i.e. $\mu \in [-\Delta, \Delta]$. 

We find that our main conclusions remain valid even when the Fermi level is not precisely tuned to the Van Hove singularity. The enhancement of the effective coupling constants with respect to the constant DOS scenario continues to exist, we can still access a ``large $K$'' regime where the non-magnetic scattering potential controls the critical $J$, and the slope of the $E_{\mathrm{S}}(J)$ curves at the Shiba transition remains independent of the impurity parameters. Further, we note that the relative change of the meaningful observables, namely $J_{\mathrm{c}}$ and the slope at the transition point, are of the order of few percent for our range of $\mu$.

In the following we present the results for the two broad classes of DOS singularities.

\subsection{Conventional Van Hove singularity}

\begin{figure}[H]
\centering
\includegraphics[width=\columnwidth]{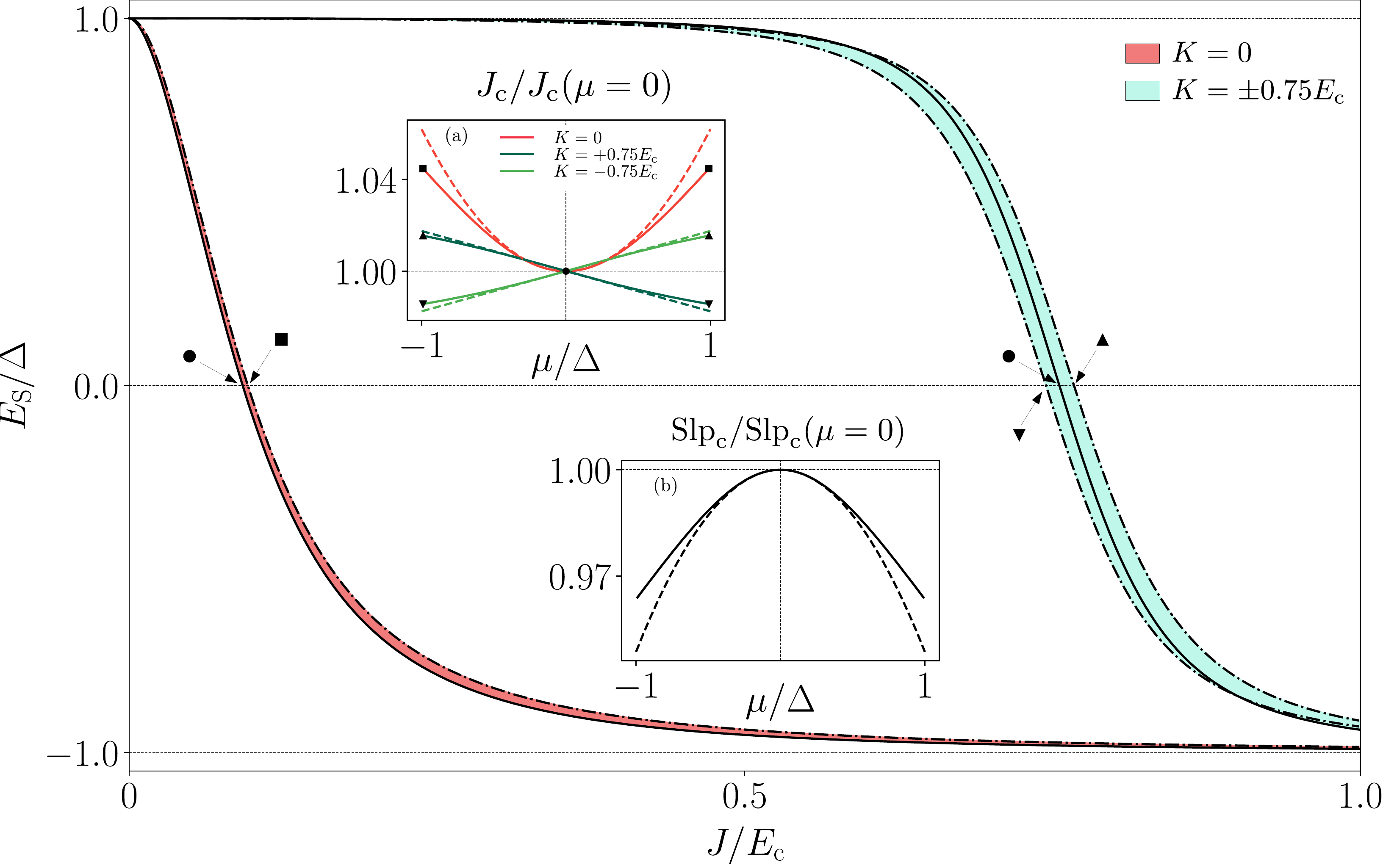}%
\caption{Variation of the positive branch of the Shiba energy, Eq.~\eqref{eq:shiba_energy_full_app} as a function of the magnetic coupling strength when the Fermi level tuned to a $[-\Delta, \Delta]$ interval around a conventional VHS. The red (green) area corresponds to a vanishing (large) non-magnetic scattering potential $K$. The solid line indicates the $\mu = 0$ curve while the dashed-dotted lines correspond to $\mu = \pm \Delta$. Inset (a) shows the normalized critical $J$ as a function of the chemical potential for different values of $K$. The markers indicate the corresponding curves in the main plot. Inset (b) depicts the normalized slope of the $E_{\mathrm{S}}(J)$ curve at the transition point as a function the chemical potential. In both insets the solid line represents the exact numerical solution and the dashed line the series expansion to lowest order in $\mu/\Delta$  [Eq.~\eqref{eq:jc_cvh_app}]. Plot parameters: $E_{\mathrm{c}} = 1000 \Delta$.}
\label{fig:mu_cvh}
\end{figure}

The Shiba energy now fulfills a self-consistent equation analogous to that discussed in Section \ref{subsec:hovh} in the context of asymmetric HOVHS:
\begin{widetext}
\begin{equation}
\label{eq:shiba_energy_full_app}
 E_{\mathrm{S}} = \pm \Delta \frac{1-2f_{2,\mathrm{vh}}K + (f_{1,\mathrm{vh}}^2+f_{2,\mathrm{vh}}^2)(K^2-J^2)}{\sqrt{4f_{1,\mathrm{vh}}^2J^2 + (1-2f_{2,\mathrm{vh}}K + (f_{1,\mathrm{vh}}^2+f_{2,\mathrm{vh}}^2)(K^2-J^2))^2}},
\end{equation}
\end{widetext}
where the $f_{2,\mathrm{vh}}$ factor stems from the asymmetry induced by the perturbation in the chemical potential. We recall that when the DOS is symmetric around the Fermi level $f_{2} = 0$ for any class of Van Hove singularities and the Shiba energy equation simplifies to the standard form [Eq.~\eqref{eq:shiba_energy}]. The renormalizing factors $f_{1,\mathrm{vh}}$ and $f_{2,\mathrm{vh}}$ are obtained by evaluating integrals \eqref{eq:I1_0_app_mu} and \eqref{eq:I2_0_app_mu}, and while it is possible to obtain a closed form in terms of the dilogarithm, the full expressions are too cumbersome to be presented here. Instead, we present the numerical solution of the self-consistent Shiba equation \eqref{eq:shiba_energy_full_app} in Fig.~\ref{fig:mu_cvh}, which clearly shows the smallness of the effect.

Nevertheless, in the limit $\frac{\mu}{\omega}\ll 1$, i.e. for a small perturbation in the chemical potential in the strong-coupling limit where $\omega^2 = \Delta^2-E_{\mathrm{S}}^2 \sim \Delta^2$, the renormalizing factors have the following approximate compact form (see App. \ref{app:integrals} for details):
\begin{subequations}
\label{eq:f1_f2_vh_app}
\begin{align}
 f_{1,\mathrm{vh}}(\Delta, E_{\mathrm{S}}, \mu) &\approx \frac{\pi}{2E_{\mathrm{c}}} \log\left(\frac{E_{\mathrm{c}}}{\sqrt{\mu^2+\omega^2}}\right),
 \label{eq:I1_0_app}
 \\
 f_{2,\mathrm{vh}}(\Delta, E_{\mathrm{S}}, \mu) &\approx \frac{\pi}{2E_{\mathrm{c}}} \frac{\mu}{\omega}.
 \label{eq:I2_0_app}
\end{align}
\end{subequations}
These expressions justify the insignificant variation of the $E_{\mathrm{S}}(J)$ curves: the factor $f_{1,\mathrm{vh}}$ inherits the logarithmic form from the DOS as in the $\mu=0$ case while the correction characterizing the newly introduced asymmetry is such that $f_{2,\mathrm{vh}}/f_{1,\mathrm{vh}} \sim \frac{\mu}{\Delta} \frac{1}{\log \left(E_{\mathrm{c}}/\Delta\right)} \ll 1$.

The approximate factors become most accurate at the Shiba transition where $E_{\mathrm{S}}$ strictly vanishes, therefore by expanding them to lowest order in $\frac{\mu}{\Delta}$ it is possible to capture the behavior of the critical magnetic coupling:
\begin{equation}
\label{eq:jc_cvh_app}
 \frac{J_{\mathrm{c}}}{J_{\mathrm{c}}(\mu = 0)} = 1 - \operatorname{sgn}(K) |\alpha_{\mathrm{vh}}| \frac{\mu}{\Delta} + |\beta_{\mathrm{vh}}| \left(\frac{\mu}{\Delta}\right)^2,
\end{equation}
where $|\alpha_{\mathrm{vh}}| \equiv |\alpha_{\mathrm{vh}}(K, \Delta, E_{\mathrm{c}})| < \frac{1}{|\log(\Delta/E_{\mathrm{c}})|}$ and $|\beta_{\mathrm{vh}}| \equiv |\beta_{\mathrm{vh}}(K, \Delta, E_{\mathrm{c}})|$, with $|\beta(K = 0)|\sim \frac{1}{|\log(\Delta/E_{\mathrm{c}})|}$. Thus, it becomes clear that the relative corrections to $J_{\mathrm{c}}$ are small assuming a realistic bandwidth, i.e. $\Delta \ll E_{\mathrm{c}}$. Interestingly, in the absence of a non-magnetic potential scattering ($K=0$) the system is particle-hole symmetric, and therefore electron or hole doping has the same effect on $J_{\mathrm{c}}$ [inset (a) in Fig.~\ref{fig:mu_cvh}]. In a likewise manner, the slope of the $E_{\mathrm{S}}(J)$ curve at the transition point is independent of the impurity parameters, and hence an even function of $\mu$ as well [inset (b) in Fig.~\ref{fig:mu_cvh}]:
\begin{align}
 \nonumber
 \mathrm{Slp}_{\mathrm{c}} \equiv \dv{E_{\mathrm{S}}}{J}\Bigr|_{J=J_{\mathrm{c}}} &= - \Delta \frac{f_{1,\mathrm{vh}}(\Delta, 0,\mu)^2 + f_{2,\mathrm{vh}}(\Delta, 0,\mu)^2}{f_{1,\mathrm{vh}}(\Delta,0,\mu)},\\ 
 \frac{\mathrm{Slp}_{\mathrm{c}}}{\mathrm{Slp}_{\mathrm{c}}(\mu=0)}&\approx 1 - |\gamma_{\mathrm{vh}}|\left(\frac{\mu}{\Delta}\right)^2,
\end{align}
where $|\gamma_{\mathrm{vh}}| \equiv |\gamma_{\mathrm{vh}} (\Delta, E_{\mathrm{c}})| \sim \frac{1}{|\log(\Delta/E_{\mathrm{c}})|}$.

\subsection{Higher order Van Hove singularity}

\begin{figure}
\centering
\includegraphics[width=\columnwidth]{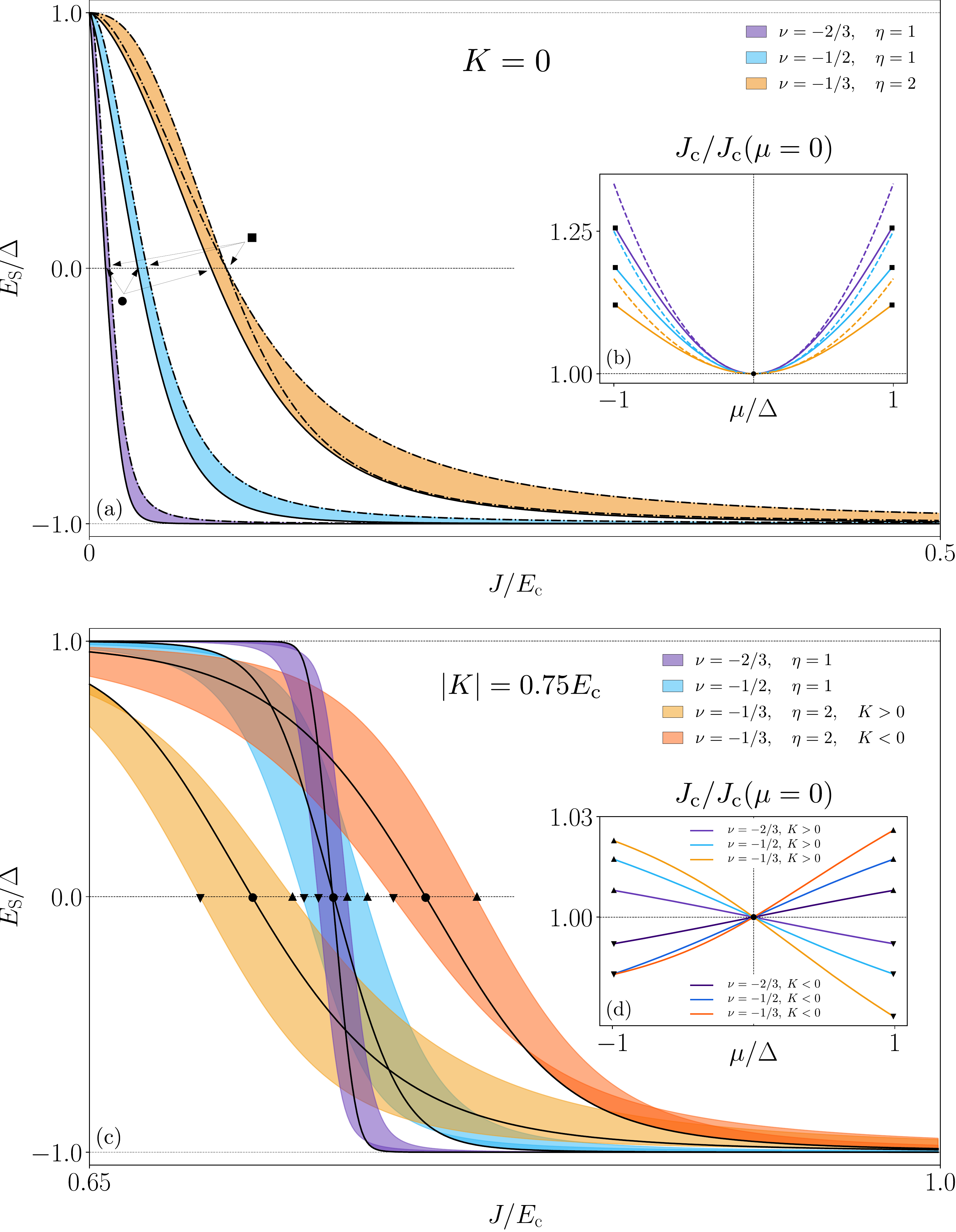}%
\caption{Variation of the positive branch of the Shiba energy as a function of the magnetic coupling strength when the Fermi level tuned to a $[-\Delta, \Delta]$ interval around different HOVHS as indicated in the legend. The solid lines indicate the $\mu = 0$ curves and the dashed-dotted lines correspond to $\mu = \pm \Delta$ (omitted in the bottom panel to improve readability). Top panel: Zero non-magnetic scattering potential. Bottom panel: Large non-magnetic scattering potential. Note that an asymmetric DOS ($\eta \neq 1$, in orange) leads to different curves for $\pm K$ but whose slope at $E_{\mathrm{S}}=0$ is the same. Insets show the normalized critical $J$ as a function of the chemical potential for different HOVHS. The relative change in $J_{\mathrm{c}}$ at $K = 0$ [inset (b)] is one order magnitude larger than in the other considered scenarios because the actual values of $J_{\mathrm{c}}$ are close to 0. The markers indicate the corresponding curves in the main plot. The dashed lines in (b) correspond to the series expansion to lowest order [Eq.~\eqref{eq:jc_hovh_app}] which was omitted in (d) to improve readability. Note that the $\pm K$ curves in the inset are symmetric upon electron or hole-doping only if the DOS is symmetric around the singularity ($\eta = 1$). Further, relative change of the absolute value of $J_{\mathrm{c}}$ is symmetric around $\mu = 0$ only if the DOS is symmetric around the singularity. Plot parameters: $E_{\mathrm{c}} = 1000 \Delta$.}
\label{fig:mu_hovh}
\end{figure}

Analogously, if the Fermi level is tuned to the vicinity of a HOVHS, the self-consistent Shiba energy equation also takes the form of Eq.~\eqref{eq:shiba_energy_full}, where the renormalization parameters now read
\begin{widetext}
 \begin{subequations}
 \label{eq:f1_f2_hovh_app}
 \begin{align}
  f'_{1,\nu}(\Delta, E_{\mathrm{S}},\mu)=&\frac{\pi}{2E_{\mathrm{c}}} \frac{1+\nu}{\cos\left(\frac{\pi}{2}\nu\right)}\left(\frac{\sqrt{\mu^2+\omega^2}}{E_{\mathrm{c}}}\right)^\nu \frac{\sin[(\pi-\varphi)\nu]+\eta \sin(\varphi \nu)}{(1+\eta)\sin\left(\frac{\pi}{2}\nu\right)},
  \\
  \nonumber
  f'_{2,\nu}(\Delta, E_{\mathrm{S}},\mu)=&\frac{\pi}{2E_{\mathrm{c}}} \frac{1+\nu}{\sin\left(\frac{\pi}{2}\nu\right)}\left(\frac{\sqrt{\mu^2+\omega^2}}{E_{\mathrm{c}}}\right)^\nu \frac{\sqrt{\mu^2+\omega^2}}{\omega}\frac{\sin[(\pi-\varphi)(\nu+1)]-\eta \sin[\varphi (\nu+1)]}{(1+\eta)\cos\left(\frac{\pi}{2}\nu\right)}\\
  &+ \frac{\mu}{\omega} f'_{1,\nu}(\Delta, E_{\mathrm{S}},\mu),
  \end{align}
 \end{subequations}
\end{widetext}
with $\varphi = \arg(\mu + i \omega)$. We find that the renormalizing parameters also inherit the power-law dependence from the DOS [cf. $\mu = 0$ case,  Eq.~\eqref{eq:f1_f2_hovh}], with $\mu$ entering as a correction to $\omega$ as it was the case in the conventional VHS. Therefore, the energy curves when the Fermi level is tuned to the vicinity of a HOVHS maintain the same structure that their $\mu=0$ counterparts [see Fig. \ref{fig:mu_hovh}]. At the Shiba transition $\omega = \Delta$ by definition, therefore a series expansion in powers of $\frac{\mu}{\Delta}$ captures the behavior of the critical $J$ at small doping. We find
\begin{equation}
\label{eq:jc_hovh_app}
 \frac{J_{\mathrm{c}}}{J_{\mathrm{c}}(\mu = 0)} = 1 - \operatorname{sgn}(K)|\alpha_{\mathrm{hovh}}| \frac{\mu}{\Delta} + \beta_{\mathrm{hovh}} \left(\frac{\mu}{\Delta}\right)^2,
\end{equation}
where $|\alpha_{\mathrm{hovh}}| \equiv |\alpha_{\mathrm{hovh}}(K, \Delta, \nu, \eta, E_{\mathrm{c}})|$ and $\beta_{\mathrm{hovh}} \equiv \beta_{\mathrm{vh}}(K, \Delta, \nu, \eta, E_{\mathrm{c}})$, with $\beta(K = 0)= -\frac{\nu}{2}$. As discussed in the previous Section, in the absence of non-magnetic scattering potential we obtain an even function of $\mu$, which interestingly only depends on the power-law exponent to lowest order. Further, for a finite $K$ the linear term does not vanish, and its sign depends on $K$ in the same manner as in the conventional VHS.

\begin{figure}[H]
\centering
\includegraphics[width=\columnwidth]{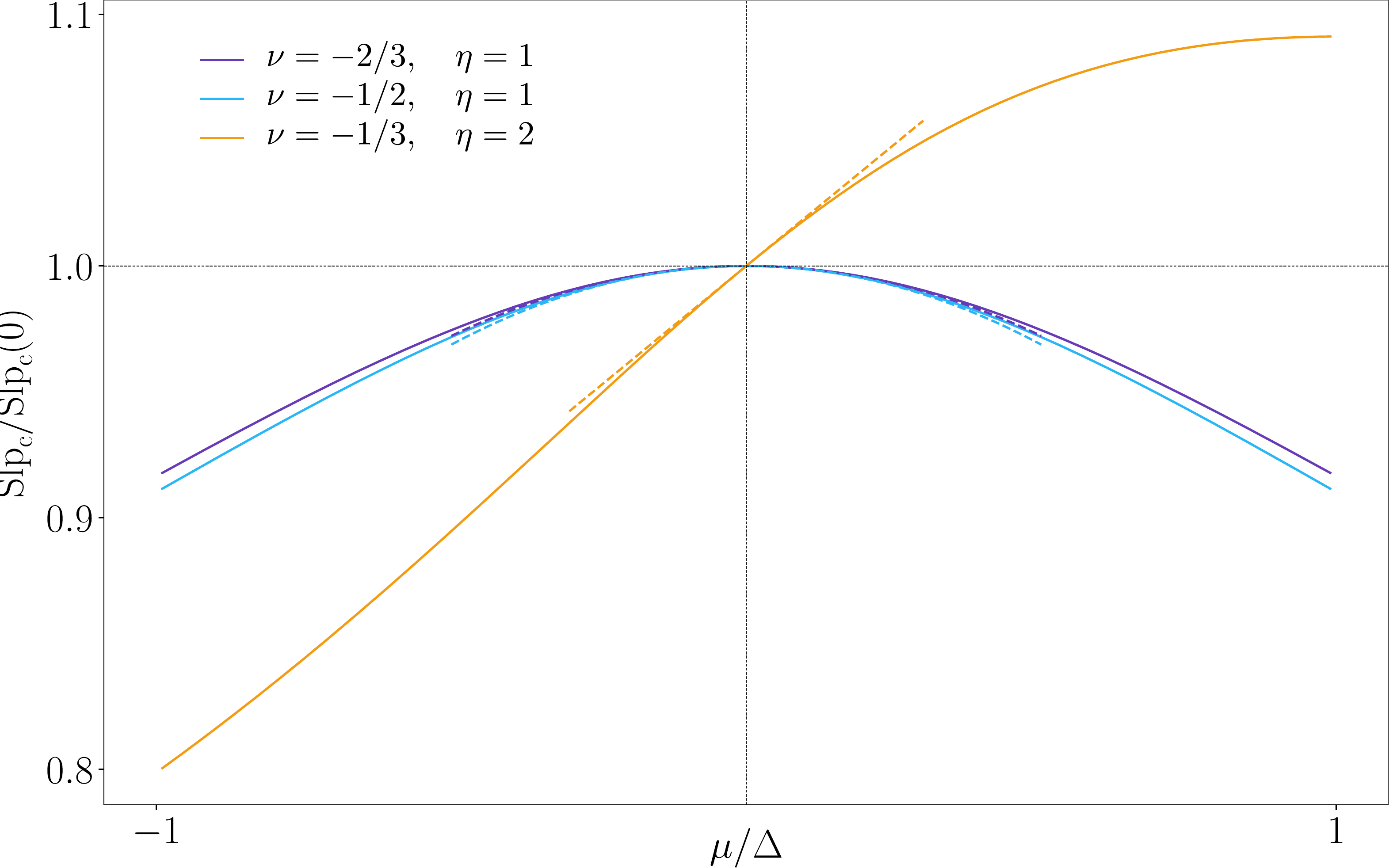}%
\caption{Normalized slope of the $E_{\mathrm{S}}(J)$ curve at the transition point as a function the chemical potential for the HOVHS presented in Fig.~\ref{fig:mu_hovh}. The solid line represents the exact solution using the renormalization parameters in Eq.~\eqref{eq:f1_f2_hovh_app} and dashed curve the series expansion to lowest order in $\mu/\Delta$ [Eq.~\eqref{eq:slope_hovh_app}]. Electron or hole doping has the same effect on the slope only if the DOS is symmetric around the singularity. Plot parameters: $E_{\mathrm{c}} = 1000 \Delta$.}
\label{fig:mu_hovh_slope}
\end{figure}

The slope of the $E_{\mathrm{S}}(J)$ curve at the Shiba transition remains independent of the impurity parameters [compare $\pm K$ curves at the same $\mu$ in Fig.~\ref{fig:mu_hovh} (c)] and it takes the form of Eq.~\eqref{eq:slope_hovh} with the newly-introduced renormalization parameters [Eq.~\eqref{eq:f1_f2_hovh_app}]. At low doping,
\begin{equation}
\label{eq:slope_hovh_app}
 \frac{\mathrm{Slp}_{\mathrm{c}}}{\mathrm{Slp}_{\mathrm{c}}(\mu=0)} \approx 1 + \gamma_{\mathrm{hovh}}\frac{\mu}{\Delta} + \delta_{\mathrm{hovh}}\left(\frac{\mu}{\Delta}\right)^2,
\end{equation}
where $\gamma_{\mathrm{hovh}} \equiv \gamma_{\mathrm{hovh}}(\nu, \eta) = \frac{\eta-1}{\eta+1}\nu \frac{\cos(\nu \pi/2)}{\sin(\nu \pi/2)}$ and $\delta_{\mathrm{hovh}} \equiv \delta_{\mathrm{hovh}}(\nu, \eta, \Delta, E_{\mathrm{c}})$, with $\gamma_{\mathrm{hovh}}(\nu, \eta = 1, \Delta, E_{\mathrm{c}}) = \nu (1 + \nu)$. The slope at the transition is only sensitive to the bulk parameters, therefore, if the DOS is symmetric around the singularity ($\eta = 1$), we obtain an even function in $\mu$ as discussed in the previous Subsection (see Fig.~\ref{fig:mu_hovh_slope}).

\section{Calculation details}
\label{app:integrals}
\subsection{Fermi level precisely tuned to the Van Hove singularity}
To calculate the energy of the in-gap YSR states $E_{\mathrm{S}}$ we introduce the spinors $\phi_{+} = (\psi_{\uparrow}, \psi^{\dagger}_{\downarrow})^T$ and $\phi_{-} = (\psi_{\downarrow}, -\psi^{\dagger}_{\uparrow})^T$ \cite{pientka:topo}. We choose the impurity to be at the origin, $\bm{r}_{\mathrm{imp}} = \bm{0}$, and we write the eigenvalue equation restricted to particle-hole space,
\begin{equation}
\label{eq:schro_eq}
(\xi_{\bm{k}} \tau_z + \Delta \tau_x - E_{\mathrm{S}}) \phi_{\pm}(\bm{k}) = - (K \tau_z \mp J \sigma_z) \phi_{\pm}(\bm{0}).
\end{equation}

The energy of the in-gap Shiba states $E_{\mathrm{S}}$ is obtained by evaluating  the Fourier transform of \eqref{eq:schro_eq} at $\bm{r} = \bm{0}$ and solving 
\begin{equation}
\label{eq:det_eq}
 \det \left[1 - \int \frac{d\bm{k}}{(2\pi)^2} \frac{E_{\mathrm{S}}+\xi_{\bm{k}}\tau_z+\Delta \tau_x}{E_{\mathrm{S}}^2-\xi_{\bm{k}}^2-\Delta^2} (K\tau_z \mp J)\right] = 0.
\end{equation}
 We solve Eq.~\eqref{eq:det_eq} by performing the substitution $\int \frac{d\bm{k}}{(2\pi)^2} = \int \rho(\xi) d\xi$ and evaluating
\begin{subequations}
\label{eq:I1_I2_0_app}
\begin{align}
 I_1(\bm{0}) &= - \int_{-E_\mathrm{c}}^{E_\mathrm{c}} d\xi \; \rho(\xi) \frac{1}{\xi^2+\omega^2},
 \label{eq:I1_0_app}
\\
 I_2(\bm{0}) &= - \int_{-E_\mathrm{c}}^{E_\mathrm{c}} d\xi \; \rho(\xi) \frac{\xi}{\xi^2+\omega^2},
 \label{eq:I2_0_app}
\end{align}
\end{subequations}
where $\omega^2 = \Delta^2-E_{\mathrm{S}}^2$.\\

Note that the same Eq.~\eqref{eq:det_eq} also follows from calculating the poles of the transfer matrix in the usual $T$-matrix method.
\subsubsection{Logarithmic DOS (conventional Van Hove singularity)}
Here
\begin{equation}
\label{eq:dos_vh_app}
\rho(\xi) = \frac{1}{2E_{\mathrm{c}}} \log\left(\left|\frac{E_\mathrm{c}}{\xi}\right|\right).
\end{equation}
In the limit $\frac{\omega}{E_{\mathrm{c}}} \rightarrow 0$,
\begin{align}
 I_1(\bm{0}) &= \frac{1}{E_{\mathrm{c}}^2} \int_0^1 \frac{dx \log(x)}{x^2+(\omega/E_{\mathrm{c}})^2}
 \nonumber
 \\
 &\sim \frac{1}{E_{\mathrm{c}}^2} \int_0^{\infty} \frac{dx \log(x)}{x^2+(\omega/E_{\mathrm{c}})^2} = -\frac{1}{2E_{\mathrm{c}}} \frac{\pi}{\omega} \log\frac{E_{\mathrm{c}}}{\omega},\\
  I_2(\bm{0}) &= \frac{1}{2E_{\mathrm{c}}} \int_{-1}^1 dx \; \log(|x|)\frac{x}{x^2+(\omega/E_{\mathrm{c}})^2}=0.
\end{align}
\subsubsection{Power-law DOS (higher order Van Hove singularity)}
Here 
\begin{equation}
\label{eq:dos_hovh_app}
\rho(\xi) = \frac{1}{1+\eta}\frac{\nu+1}{E_{\mathrm{c}}^{1+\nu}} \; |\xi|^{\nu} \begin{cases} 
      \eta \quad \mathrm{if} \quad \xi < 0, \\
      1 \quad \mathrm{if} \quad \xi > 0.\end{cases}
\end{equation}
with $-1 < \nu < 0$, $\nu \in \mathbb{Q}$ . To calculate the integrals we use result 3.251.10 from Ref. \cite{gradshteyn2007}:
\begin{equation}
 \int_0^1 dx \; x^{p-1} (1-x^q)^{-\frac{p}{q}} = \frac{\pi}{q}\frac{1}{\sin(\frac{p}{q} \pi)},  \quad q>p>0.
\end{equation}
Taking an infinite energy cutoff, $E_{\mathrm{c}} \rightarrow \infty$, we have
\begin{widetext}
\begin{subequations}
\begin{align}
 I_1(\bm{0}) &= - \frac{\nu+1}{E_{\mathrm{c}}^{1+\nu}}  \omega^{\nu-1} \int_0^{\infty} dx \; \frac{x^{\nu}}{x^2+1}=  - \frac{\nu+1}{2E_{\mathrm{c}}^{1+\nu}} \omega^{\nu-1} \int_0^1 dt \; t^{-\frac{1+\nu}{2}} (1-t)^{\frac{\nu-1}{2}}=-\pi \frac{\nu+1}{2E_{\mathrm{c}}^{1+\nu}\cos\left(\frac{\pi}{2}\nu\right)} \omega^{\nu-1},\label{eq:calc_I1_0}\\ 
 I_2(\bm{0}) &=  -\frac{\nu+1}{E_{\mathrm{c}}^{1+\nu}} \frac{1-\eta}{1+\eta}  \omega^{\nu} \int_0^{\infty} dx \; \frac{x^{\nu+1}}{x^2+1}= \frac{\nu+1}{2E_{\mathrm{c}}^{1+\nu}} \frac{\eta-1}{1+\eta}\omega^{\nu} \int_0^1 dt \; t^{-1-\frac{\nu}{2}} (1-t)^{\frac{\nu}{2}} \label{eq:calc_I2_0} =\pi \frac{1-\eta}{1+\eta} \frac{\nu+1}{2E_{\mathrm{c}}^{1+\nu}\sin\left(\frac{\pi}{2}\nu\right)} \omega^{\nu},
\end{align}
\end{subequations}
\end{widetext}
where we performed the changes of variables $x=\frac{\xi}{\omega}$ and $t = \frac{1}{x^2+1}$, and we identified $q=1$ and $p = \frac{1-\nu}{2}$ \eqref{eq:calc_I1_0} and $p = -\frac{\nu}{2}$ \eqref{eq:calc_I2_0}.
\subsection{Fermi level tuned to an interval $[-\Delta, \Delta]$ around the Van Hove singularity}
We now allow for a finite chemical potential and the relevant integrals read

\begin{subequations}
\label{eq:I1_I2_0_app_mu}
\begin{align}
 I_1^\mu(\bm{0}) &= - \int_{-E_\mathrm{c}}^{E_\mathrm{c}} d\xi \; \rho(\xi) \frac{1}{(\xi-\mu)^2+\omega^2},
 \label{eq:I1_0_app_mu}
\\
 I_2^\mu(\bm{0}) &= - \int_{-E_\mathrm{c}}^{E_\mathrm{c}} d\xi \; \rho(\xi) \frac{\xi-\mu}{(\xi-\mu)^2+\omega^2},
 \label{eq:I2_0_app_mu}
\end{align}
\end{subequations}
where $\omega^2 = \Delta^2-E_{\mathrm{S}}^2$.\\

\subsubsection{Logarithmic DOS (conventional Van Hove singularity}
Substituting Eq.~\eqref{eq:dos_vh_app} and in the limit $\frac{\omega}{E_{\mathrm{c}}} \rightarrow 0$,
\begin{widetext}
\begin{align}
 I_1^\mu(\bm{0}) &\sim \frac{1}{2E_{\mathrm{c}}^2} \int_0^\infty dx \;\left\{\frac{\log(x)}{[x-(\mu/E_{\mathrm{c}})]^2+(\omega/E_{\mathrm{c}})^2} +\frac{\log(x)}{[x+(\mu/E_{\mathrm{c}})]^2+(\omega/E_{\mathrm{c}})^2}\right\} = \frac{-\pi}{2E_{\mathrm{c}} \omega} \log\left(\frac{E_{\mathrm{c}}}{\sqrt{\mu^2+\omega^2}}\right),
 \label{eq:I1_0_app_mu_sol}
 \\
  I_2^\mu(\bm{0}) &\sim -\frac{1}{2E_{\mathrm{c}}} \int_0^\infty dx \; x\log(x)\left\{\frac{1}{[x+(\mu/E_{\mathrm{c}})]^2+(\omega/E_{\mathrm{c}})^2} - \frac{1}{[x+(\mu/E_{\mathrm{c}})]^2+(\omega/E_{\mathrm{c}})^2} \right\} - \mu \; I_1^\mu(\bm{0}) \sim \frac{\pi}{2E_{\mathrm{c}}} \frac{\mu}{\omega},
\label{eq:I2_0_app_mu_sol}
\end{align}
 \end{widetext}
 
 where in integral \eqref{eq:I2_0_app_mu_sol} we further took the limit $\mu \ll \omega$ and approximated the term in brackets by $-\frac{\mu}{E_{\mathrm{c}}}\frac{4x}{(x^2+(\omega/E_{\mathrm{c}})^2)^2}$.
 
\subsubsection{Power-law DOS (higher order Van Hove singularity)}
Substituting Eq.~\eqref{eq:dos_hovh_app}, the integrals \eqref{eq:I1_I2_0_app_mu} are readily solved using the standard keyhole contour of infinite radius with a branch cut on the positive real axis:
\begin{widetext}
\begin{align}
I_1^\mu(\bm{0}) &\sim -\pi (1+\nu) \frac{R^\nu}{E_{\mathrm{c}}^{1+\nu}\omega}\frac{\sin[(\pi-\varphi)\nu]+\eta \sin(\varphi\nu)}{(1+\eta)\sin(\pi \nu)},
\\
I_2^\mu(\bm{0}) &\sim -\pi (1+\nu) \frac{R^{1+\nu}}{E_{\mathrm{c}}^{1+\nu}\omega}\frac{\sin[(\pi-\varphi)(\nu+1)]-\eta \sin[\varphi(\nu+1)]}{(1+\eta)\sin[\pi (\nu+1)]} - \mu \; I_1^\mu(\bm{0}),
\end{align}
\end{widetext}
where $R e^{i \varphi} = \mu + i \omega$.

\begin{acknowledgments}
We would like to thank Tristan Cren for fruitful discussions.
\end{acknowledgments}

\bibliography{bibliography.bib}

\end{document}